\documentstyle[aps, prabib, tighten]{revtex}

\begin{document}

% \jl{3}

%%-> 
\draft

\title{High-field noise in metallic diffusive conductors}

\author{Frederick Green$\dagger$
and
Mukunda P Das$\ddagger$}

\address{
$\dagger$
GaAs IC Prototyping Facility,
CSIRO Telecommunications and Industrial Physics,
PO Box 76, Epping NSW 1710, Australia
}

\address{
$\ddagger$
Department of Theoretical Physics,
Research School of Physical Sciences and Engineering,
The Australian National University,
Canberra ACT 0200, Australia
}

\maketitle

\begin{abstract}
We analyze high-field current fluctuations in degenerate
conductors by mapping the electronic Fermi-liquid correlations
at equilibrium to their semi-classical non-equilibrium form.
Our resulting Boltzmann description is applicable
to diffusive mesoscopic wires.
We derive a non-equilibrium connection between thermal fluctuations
of the current and resistive dissipation. In the weak-field limit this
is the canonical fluctuation-dissipation theorem.
Away from equilibrium, the connection enables explicit
calculation of the excess ``hot-electron'' contribution
to the thermal spectrum. We show that
excess thermal noise is strongly inhibited by Pauli exclusion.
This behaviour is generic to the semi-classical metallic regime.
\end{abstract}

\bigskip

%IOP+ \maketitle

\section{Introduction}

In this paper we address a technologically important open problem:
non-equilibrium noise in strongly driven degenerate conductors.
Nano-fabrication has made possible a variety of
refined measurements of transport and noise, for many different
structures at sub-micron dimensions
\cite{liefr,rez,ksgje,smd,sbkpr,poth,depic,sami}.
Alongside the experiments
there has been much theoretical activity
\cite{theor,khlus,lesovik,but,beebut,thmldr,nagaev1,djb2}.

A thriving topic is the behaviour of current fluctuations in
diffusive wires. Typically, this concerns structures
shorter than the bulk inelastic mean free path but
still much longer than that for elastic scattering.
They are in an operating region where randomness of the
carrier motion prevails. This is our regime of interest.

Two theories have come to the fore as
methods of choice for describing mesoscopic transport.
These are inherently {\it weak-field} models,
predicated upon exclusively linear forms of transport analysis.
One technique (Landauer-B\"uttiker) is
based on coherent quantum transmission
\cite{theor,blbu}.
This has been adapted to fluctuations and noise by Khlus
\cite{khlus},
Lesovik
\cite{lesovik},
Beenakker and B\"uttiker
\cite{but,beebut},
Martin and Landauer
\cite{thmldr},
and many others
\cite{theor,blbu}.
Another approach (Boltzmann-Langevin) uses
stochastic transport equations
\cite{theor},
reduced to a diffusive model
\cite{datta}.
The same phenomenology has since been
applied to fluctuations and noise by Nagaev
\cite{nagaev1}
and de Jong and Beenakker
\cite{theor,djb2}.

Although these mesoscopic-noise methodologies
are markedly distinct, both agree on their paradigm. They view
a mesoscopic wire as a random assembly of
individual elastic scatterers, in a bath of free carriers whose
propagation, impeded by the scatterers, must be regarded
as {\it strictly} diffusive and must be calculated as such
\cite{datta}.
For an exhaustive survey of diffusive noise theories
we cite the recent review of Blanter and B\"uttiker
\cite{blbu}
as well as the earlier one of de Jong and Beenakker
\cite{theor}.

The leading diffusive methods for noise also
share a number of difficulties.
Not least among these is the issue of conformity with the
fluctuation structure of charged Fermi liquids
\cite{csr}.
This and other basic problems are analyzed in detail in references
\onlinecite{upon99} and \onlinecite{ithaca}.

Clean, truly first-principles solutions
certainly exist for non-equilibrium noise.
Monte Carlo simulations are very well established,
as witness those of the Lecce group
\cite{lecce1},
even if high-field Monte Carlo is still rare for metals
\cite{mc,lecce2}.
In the non-degenerate case an analytical, self-contained and
computable theory of noise has been formulated
by Korman and Mayergoyz
\cite{kormay}.
Their approach is strictly kinetic and free
of superfluous phenomenological props. In philosophy
it is akin to Green-function models for fluctuations,
such as Stanton and Wilkins'
\cite{sw2,sw0}.

In developing a microscopically consistent account of noise,
there are cogent reasons to stay within the traditional
kinetic canons rather than embrace novel diffusive doctrines.
The chief reason, perhaps, is sheer technological need;
device designers can scarcely afford to be hobbled
by phenomenologies whose congenital linearity
denies any access to the vital high-field region.
The small scale of modern device structures means that
they are routinely driven into non-linear response
\cite{ferry}.

To illustrate this point we estimate the range of validity
for linear diffusion.
In the weak-field limit the Einstein relation
\cite{datta,upon99}, or drift-diffusion equivalence,
underpins diffusive transport. Roughly speaking,
drift-diffusion equivalence breaks down (and linear
diffusive transport with it) when the energy gained
in drift mediated by inelastic scattering
exceeds the energy scale for diffusion
mediated by elastic processes.
In a short metallic wire this means that
$(eV/L)\min{\{L, L_{\rm in}\}} \gtrsim \hbar v_{\rm F}/L_{\rm el}$,
where $V$ is the driving voltage, $L$ is the sample length
and $L_{\rm in}$ is the (bulk) inelastic mean free path
at the Fermi surface. Similarly $L_{\rm el}$ is
the elastic mean free path ($v_{\rm F}$ is the Fermi velocity).
Note that for $L_{\rm in} > L$
the effective inelastic path becomes the sample length,
since dissipation in the bounding leads is dominant.

For a typical mesoscopic silver wire of electron
density $6\!\times\!10^{22}{~}{\rm cm}^{-3}$,
the low-temperature transport parameters are
\cite{smd}
$L_{\rm in} = 1 {~}{\rm cm} \gg L = 30{~}\mu$m and
$L_{\rm el} = 50{~}$nm.
The threshold voltage is then $V \gtrsim 40{~}$mV
for the breakdown of the diffusive regime.
In a two-dimensional mesoscopic channel
at density $2\!\times\!10^{11}{~}{\rm cm}^{-2}$
the threshold is appreciably lower, with
\cite{liefr}
$L = 17{~}\mu$m, $L_{\rm in} = 6{~}\mu$m and
$L_{\rm el} = 1.4{~}\mu$m
giving $V \gtrsim 0.25{~}$mV.
This shows how readily mesoscopic devices,
particularly low-dimensional ones,
can enter the non-linear regime beyond diffusive theory.

The present is the first of three studies covering
the essential formalism for non-equilibrium noise,
the action of Coulomb correlations in non-uniform systems
\cite{gdii}
and finally the kinetic description of shot noise
\cite{gdcond}.
Throughout, we follow one overarching principle. It is that
a consistent model of non-equilibrium fluctuations will conform
to basic properties of the electron gas in a natural way,
if and {\it only} if such a model is grounded explicitly in the
theory of the electronic Fermi liquid.

To require that a kinetic description of fluctuations and noise
respect fundamental conservation laws in equilibrium, is to place
a unifying constraint on its low- and high-field forms together.
Our aim is to catalogue all the physical consequences of this assertion.
A viable kinetic model will necessarily recover
the fluctuation-dissipation relation
\cite{kogancpu},
but it must also contain the equally fundamental conserving sum rules
\cite{pinoz}.

Much of the authoritative literature on diffusive noise theory,
if not all of it, prefers to make a virtue
of its heavy dependence on drift-diffusion
equivalence and the fluctuation-dissipation theorem,
while remaining oblivious to every other
basic sum-rule requirement. Such an understanding is too scanty.
It is the {\it full} physics
of Fermi liquids which governs their fluctuations,
a fact which cannot simply be ignored.

An established tool for scattering-dominated
noise in degenerate conductors
is the semi-classical Green-function approach
\cite{sw2,sw0,kogan,ggk,gc}.
We take it beyond its well-understood role in time-dependent
response, by proving that the dynamical Green function
also governs the structure of the adiabatic (steady-state)
fluctuations. In turn, these determine the mean initial
strength of the time-dependent fluctuations.

The non-equilibrium adiabatic fluctuations are exact
closed functionals of their equilibrium form.
This offers the key to practical and flexible
calculations over a wide range of useful non-perturbative
collision models. (Here the detailed classical studies
of Stanton and Wilkins
\cite{sw2,sw0}
show the way.)
Such problems are entirely out of range for the
Boltzmann-Langevin models,
of wide currency but quite narrow practicality
for conductors in strong driving fields
\cite{sw2},
with strong internal interactions
\cite{nvk}.

In section 2 we present a wholly conventional
Boltzmann description of carrier fluctuations
in non-uniform metallic systems,
down to the {\it same} mesoscopic range accessible
to alternative (diffusive) models
\cite{theor,blbu}.
We demonstrate the quantitative connection
between fluctuations and power dissipation
well out of equilibrium. In the weak-field
limit, this connection is the canonical
fluctuation-dissipation theorem (FDT). At high fields, it
describes the hot-electron contribution to current noise.
This lets us calculate, in section 3,
the excess thermal spectrum, which is strongly
suppressed in a degenerate system.
In section 4 we sum up, and preview the two forthcoming works.

\section{Theory}

The theoretical discussion is in six parts. We begin
by formulating the transport problem as a direct mapping of the
electron Fermi liquid to its non-equilibrium steady state.
Next we describe the steady-state fluctuations,
after which we discuss time dependence, then the
dynamic fluctuations and their formal
connection with the steady state.
This produces a self-contained expression for the
current-current fluctuation, which determines thermal noise.
Last, we analyze the connection between fluctuations
and dissipation in the non-equilibrium region.

\subsection{Transport Model}

The semi-classical Boltzmann transport
equation for the electron distribution function
$f_{\alpha} (t) \equiv f_{s}({\bf r}, {\bf k}, t)$ is

\begin{eqnarray}
%IOP+ \fl
{\left[
{ {\partial}\over {\partial t} } +
  { {\bf v}_{{\bf k} s } }{\bbox \cdot}
 { {\partial}\over {\partial {\bf r} }} -
  {
   { { e{\bf E}({\bf r},t) }\over {\hbar} }{\bbox \cdot}
  }
   { {\partial}\over {\partial {\bf k}} }
\right]} f_{\alpha}(t)
=
&& 
-\sum_{\alpha'}
{\Bigl[
W_{\alpha' \alpha} (1 - f_{\alpha'}) f_{\alpha}
%\Bigr.} \cr
%&& {\Bigl.
- W_{\alpha \alpha'} (1 - f_{\alpha}) f_{\alpha'}
\Bigr]}.
\label{AX1}
\end{eqnarray}

\noindent
Label $\alpha = \{{\bf k}, s, {\bf r}\}$ denotes a
point in single-particle phase space, while sub-label {\it s}
indexes both the discrete sub-bands (or valleys) of
a multi-level system and the spin state.
The system is acted upon by the total
field ${\bf E}({\bf r},t)$.
We study single-particle scattering, with a rate
$W_{\alpha \alpha'} \equiv
\delta({\bf r} - {\bf r'})
W_{s s'}({\bf k}, {\bf k'}; {\bf r})$ that is local in real space,
independent of the driving field, and that satisfies detailed balance:
$W_{\alpha' \alpha} (1 - f^{\rm eq}_{\alpha'}) f^{\rm eq}_{\alpha}
= W_{\alpha \alpha'} (1 - f^{\rm eq}_{\alpha}) f^{\rm eq}_{\alpha'}$
where $f^{\rm eq}_{\alpha}$ is the equilibrium distribution.
In a system with $\nu$ dimensions, we make
the following correspondence for the identity operator:

\[
\delta_{\alpha \alpha'}
\equiv \delta_{s  s'}
{\left\{ { {\delta_{{\bf r} {\bf r'}}}\over
	   {\Omega({\bf r})} } \right\}}
{\left\{ \Omega({\bf r}) \delta_{{\bf k} {\bf k'}} \right\}}
\longleftrightarrow \delta_{s  s'}
\delta({\bf r} - {\bf r'})
(2\pi)^{\nu} \delta({\bf k} - {\bf k'}).
\]

\noindent
The volume $\Omega({\bf r})$
of a local cell in real space becomes the measure for
spatial integration, while its inverse defines the
scaling in wave-vector space
for the local bands $\{ {\bf k}, s \}$.

The first step is to construct the steady-state solution
$f_{\alpha} \equiv f_{\alpha}(t \to \infty)$ explicitly
from $f^{\rm eq}$, which satisfies the equilibrium
form of equation (\ref{AX1}):

\begin{eqnarray}
%IOP+ \fl
{\left[
  { {\bf v}_{{\bf k} s } }{\bbox \cdot}
 { {\partial}\over {\partial {\bf r} }} -
  {
   { { e{\bf E}_0({\bf r}) }\over {\hbar} }{\bbox \cdot}
  }
   { {\partial}\over {\partial {\bf k}} }
\right]} f^{\rm eq}_{\alpha}
= 0 =
%%% && 
-\sum_{\alpha'}
{\Bigl[
W_{\alpha' \alpha} (1 - f^{\rm eq}_{\alpha'}) f^{\rm eq}_{\alpha}
%%% \Bigr.} \cr
%%% && {\Bigl.
- W_{\alpha \alpha'} (1 - f^{\rm eq}_{\alpha}) f^{\rm eq}_{\alpha'}
\Bigr]}.
\label{AX1.1}
\end{eqnarray}

\noindent
The internal field ${\bf E}_0({\bf r})$ is defined in the absence of a
driving field. The quantities $f^{\rm eq}$ and ${\bf E}_0$
are linked self-consistently by the usual constitutive relations,
the first being the Poisson equation

\begin{mathletters}
\label{AX1.2}

\begin{equation}
{\partial\over {\partial {\bf r}} } {\bbox \cdot}
\epsilon {\bf E}_0 = -4\pi e
{\Bigl(  {\langle f^{\rm eq}({\bf r}) \rangle} - n^+({\bf r}) \Bigr)}
\label{AX1.2a}
%\label{poissoneq}
\end{equation}

\noindent
in terms of the dielectric constant $\epsilon({\bf r})$,
the electron density
$\langle f^{\rm eq}({\bf r}) \rangle \equiv
{\Omega({\bf r})}^{-1}{\sum}_{{\bf k},s}
f^{\rm eq}_{\alpha}$,
and the positive background density $n^+({\bf r})$,
which is taken to be independent of the driving field
\cite{embed}.
Normalization to the total particle number is
$\sum_{\bf r} \Omega({\bf r}) {\langle f^{\rm eq}({\bf r}) \rangle} = N$.
The second relation is the form of the equilibrium function itself,

\begin{equation}
f^{\rm eq}_{\alpha} ~=~
{\left[
1 + \exp \!
 {\left(
   { {\varepsilon_{\alpha} - \phi_{\alpha}}\over k_{B}T }
  \right)}
\right]}^{-1}
\label{AX1.2b}
\end{equation}

\end{mathletters}

\noindent
at temperature $T$. The conduction-band energy
$\varepsilon_{\alpha} = \varepsilon_s({\bf k}; {\bf r})$
may have structural parameters that depend on position implicitly.
The locally defined Fermi level
$\phi_{\alpha} = \mu - V_0({\bf r})$ is the difference
of the global chemical potential $\mu$ and the
electrostatic potential $V_0({\bf r})$,
whose gradient is $e{\bf E}_0({\bf r})$.

Define the difference function 
$g_{\alpha} = f_{\alpha} - f^{\rm eq}_{\alpha}$.
From each side of equation (\ref{AX1}) in the steady state,
subtract its equilibrium counterpart
\cite{variat}.
We obtain

\begin{eqnarray}
%IOP+ \fl
  { {\bf v}_{{\bf k} s } }{\bbox \cdot} 
 {{\partial g_{\alpha} }\over { \partial {\bf r} }} -
  {
   { { e{\bf E}({\bf r}) }\over {\hbar} }{\bbox \cdot}
  }
   { {\partial g_{\alpha} }\over {\partial {\bf k}} } =&&
  {
   { { e({\bf E} - {\bf E}_0) }\over {\hbar} }{\bbox \cdot}
  }
   { {\partial f^{\rm eq}_{\alpha}
     }\over {\partial {\bf k}} }
- \sum_{\alpha'}
( W_{\alpha' \alpha} g_{\alpha} - W_{\alpha \alpha'} g_{\alpha'} )
\cr
&&+ \sum_{\alpha'}
( W_{\alpha' \alpha} - W_{\alpha \alpha'} )
( f^{\rm eq}_{\alpha'} g_{\alpha}
+ g_{\alpha'} f^{\rm eq}_{\alpha} + g_{\alpha} g_{\alpha'} ).
\label{AX3}
\end{eqnarray}

\noindent
The solutions to equations (\ref{AX1.1}) and
(\ref{AX3}) are determined by the
asymptotic conditions in the source and drain reservoirs,
be it at equilibrium or with an external electromotive force.
The active region includes the carriers within
source and drain terminals out to several
screening lengths. This means that local fields
are negligible at the interfaces with the reservoirs;
in practice, one shorts out the fields so that
${\bf E}({\bf r}) = {\bf E}_0({\bf r}) = {\bf 0}$
beyond these boundaries. Then Gauss's theorem implies that the
system remains globally neutral:

\begin{equation}
\sum_{\bf r} \Omega({\bf r})\langle g ({\bf r}) \rangle
\equiv \sum_{\alpha} g_{\alpha} = 0.
\label{gauss}
\end{equation}

Recast equation (\ref{AX3}) as an integro-differential equation.
The inhomogeneous term on its right-hand side generates
the explicit dependence on the equilibrium state of the system:

\begin{equation}
\sum_{\alpha'}
B[W^A f]_{\alpha \alpha'} g_{\alpha'}
= { { e{\bf {\widetilde E}} ({\bf r})}
\over {\hbar} }{\bbox \cdot}
   { {\partial  f^{\rm eq}_{\alpha}
}\over {\partial {\bf k}} }
+ \sum_{\alpha'}
W^A_{\alpha \alpha'} g_{\alpha'} g_{\alpha}.
\label{AX5}
\end{equation}

\noindent
The net non-equilibrium field ${\bf E} - {\bf E}_0$
is represented here as ${\bf {\widetilde E}}
\equiv {\bf E}_{\rm ext} + {\bf E}_{\rm ind}$,
consisting of ${\bf E}_{\rm ext}({\bf r})$,
the externally applied field
\cite{wmws},
plus the local induced response
${\bf E}_{\rm ind}({\bf r})$.
The linearized Boltzmann operator $B[W^A f]$
is

\begin{eqnarray}
%IOP+ \fl
B[W^A f]_{\alpha \alpha'}
=&&
\delta_{\alpha \alpha'}
{\left[
{ {\bf v}_{{\bf k'} s'} }{\bbox \cdot} 
 {{\partial }\over { \partial {\bf r'} }} -
  {
   { { e{\bf E}({\bf r'}) }\over {\hbar} }{\bbox \cdot}
  }
   { {\partial }\over {\partial {\bf k'}} } +
  \sum_{\beta} ( W_{\beta \alpha'}
- W^A_{\beta \alpha'} f_{\beta} )
\right]}
%\cr
%&&
- W_{\alpha \alpha'} + W^A_{\alpha \alpha'} f_{\alpha},
\label{bop}
\end{eqnarray}

\noindent
with $W^A_{\alpha \alpha'} = W_{\alpha \alpha'} - W_{\alpha' \alpha}$.
Note that $W^A = 0$ if the scattering is elastic
or if a linear approximation (such as Drude)
replaces the explicit Boltzmann collision term.

If it is to represent the physical solution,
$g$ must vanish with ${\bf {\widetilde E}}$ in the equilibrium limit.
This is guaranteed by the Poisson equation for the induced field,

%\begin{equation}
\[
{\partial\over {\partial {\bf r}} } {\bbox \cdot}
\epsilon {\bf E}_{\rm ind}
= -4\pi e
{\Bigl( \langle f({\bf r}) \rangle
- \langle f^{\rm eq}({\bf r}) \rangle \Bigr)}
= -4\pi e \langle g ({\bf r}) \rangle.
\]
%\label{poiss}
%\end{equation}

\subsection{Steady-State Response}

To calculate the adiabatic response of the system
about its steady non-equilibrium operating point,
we introduce the propagator
\cite{fg1}

\begin{equation}
G_{\alpha \alpha'}
\buildrel \rm def \over =
{ { \delta g_{\alpha} }\over
  { \delta f^{\rm eq}_{\alpha'} } },
\label{AX6.1}
\end{equation}

\noindent
with a global constraint following
directly from equation (\ref{gauss}):

\begin{equation}
\sum_{\alpha} G_{\alpha \alpha'} = 0 {~~} {\rm for {~}all} {~}\alpha'.
\label{dgauss}
\end{equation}

\noindent
The equation for $G$ is derived by taking variations
on both sides of equation (\ref{AX3}):

\begin{equation}
%IOP+ \fl
\sum_{\beta}
B[W^A f]_{\alpha \beta} G_{\beta \alpha'}
= \delta_{\alpha \alpha'}
{\left[
{ { e{\bf {\widetilde E}} ({\bf r'})}
\over {\hbar} }{\bbox \cdot}
   { {\partial }\over {\partial {\bf k'}} }
+ \sum_{\beta} W^A_{\beta \alpha'} g_{\beta}
\right]} - W^A_{\alpha \alpha'} g_{\alpha}.
\label{AXG}
\end{equation}

\noindent
The variation is restricted by excluding the reaction of the local fields
${\bf E}_0({\bf r})$ and ${\bf E}({\bf r})$.
This means that $G$ is a response function free of Coulomb screening.
Here we treat the electrons
as an effectively neutral Fermi liquid. In our second paper we will
describe the complete fluctuation structure, with Coulomb effects
\cite{gdii}.

All of the steady-state fluctuation properties
induced by the thermal background
will be specified in terms of $G$
and the equilibrium two-body fluctuation.
This consists of the ``proper'' electron-hole pair
correlation in its static long-wavelength limit
(up to a normalization factor).
In the free-electron approximation
\cite{pinoz},
that correlation is

\begin{mathletters}
\label{AX7.0}

\begin{equation}
\lim_{q \ll k_{\rm F}} {\left[
\lim_{\omega \to 0} {\left(
{{f^{\rm eq}_s({\bf r}, {\bf k}\!-\!{\bf q}/2)
- f^{\rm eq}_s({\bf r}, {\bf k}\!+\+{\bf q}/2)}
\over {\hbar\omega - \varepsilon_s({\bf k}\!+\!{\bf q}/2; {\bf r})
+ \varepsilon_s({\bf k}\!-\!{\bf q}/2; {\bf r})}}
\right)} \right]}
= -{{\delta f^{\rm eq}_{\alpha}}\over
{\delta \varepsilon_{\alpha}}}
= {{\delta f^{\rm eq}_{\alpha}}\over {\delta \phi_{\alpha}}}
\label{AX7.0a}
\end{equation}

\noindent
where the net momentum transfer becomes negligible relative
to the Fermi wave number $k_{\rm F}$.
The formal statistical-mechanical definition of
the (mean square) occupation-number {\it fluctuation}
$\Delta f^{\rm eq}$ takes it as the variation of
the occupancy $f^{\rm eq}$ of equation (\ref{AX1.2b})
with respect to the electro-chemical potential
and normalized to the thermal energy,
keeping $T$ and the local volume $\Omega({\bf r})$ fixed.
That is,

\begin{equation}
\Delta f^{\rm eq}_{\alpha} \equiv
k_{\rm B} T { {\delta f^{\rm eq}_{\alpha} }
      \over {\delta \phi_{\alpha}} }.
\label{AX7.0b}
\end{equation}

\noindent
It is easy to derive the free-electron form of the
equilibrium fluctuation:

\begin{equation}
\Delta f^{\rm eq}_{\alpha}
= f^{\rm eq}_{\alpha}(1 - f^{\rm eq}_{\alpha}).
\label{AX7.0c}
\end{equation}

\end{mathletters}

\noindent
When there are strong exchange-correlation interactions
the two-body correlation, equation (\ref{AX7.0a}),
is renormalized by a coefficient that
depends on the Landau quasi-particle parameters.
This coefficient carries through in $\Delta f^{\rm eq}$.
In the present model we neglect exchange-correlation effects.
This is a valid approximation in dense degenerate systems
\cite{pinoz}.

Before discussing the non-equilibrium fluctuations
we comment on the crucial contrast between the
quantum-Fermi-liquid origin of equation (\ref{AX7.0})
and the widespread Boltzmann-Langevin approach,
which is essentially classical.
Equation (\ref{AX7.0a}) describes an elementary
and {\it kinematically coupled} electron-hole excitation
[a fact that is particularly obvious in the particle-hole
structure of the energy denominator,
${\hbar\omega - \varepsilon_s({\bf k}\!+\!{\bf q}/2; {\bf r})
+ \varepsilon_s({\bf k}\!-\!{\bf q}/2; {\bf r})}$].
Its form is determined by the same quantum dynamical equation
whose semi-classical limit is the Boltzmann equation itself
\cite{kb}.

The physical character of a polarized excitation demands
its representation as a self-contained entity. In
thermodynamic terms, stochasticity attaches to the spontaneous
generation of electron-hole {\it pairs} in the system
and not to their electron and hole constituents independently
(as if pairwise charge balance were of no real consequence).
By its nature, a pair fluctuation cannot
be decomposed {\it ad hoc} into two stochastically
unlinked single-particle factors.
The Boltzmann-Langevin approach, on the other hand,
is tantamount to such a notional decomposition
\cite{kogancpu}.
Neither natural nor necessary for the right description
of the elementary fluctuations in a charged Fermi liquid,
it is eminently dispensable.

Define the two-point particle-hole function
$\Delta f^{(2)}_{\alpha \alpha'} \equiv
( \delta_{\alpha \alpha'} + G_{\alpha \alpha'} )
\Delta f^{\rm eq}_{\alpha'}$.
The steady-state distribution of the local number fluctuation
is the sum of all of the two-body terms:

\begin{equation}
\Delta f_{\alpha}
= \sum_{\alpha'} \Delta f^{(2)}_{\alpha \alpha'}
= \Delta f^{\rm eq}_{\alpha}
+ \sum_{\alpha'} G_{\alpha \alpha'} \Delta f^{\rm eq}_{\alpha'} {~~~}
{\rm for {~}all} {~}\alpha.
\label{AX7}
\end{equation}

\noindent
Once the explicit solution for $G$ is obtained,
the behaviour of $\Delta f$ is known completely.
This non-equilibrium fluctuation
satisfies the linearized steady-state
Boltzmann equation:

\begin{equation}
\sum_{\beta} B[W^A f]_{\alpha \beta} {\Delta f}_{\beta} = 0.
\label{AX7.B}
\end{equation}

\noindent
We see that the solution to this equation is
manifestly a linear functional of its equilibrium counterpart.
Jointly, equations (\ref{AX7}) and (\ref{AX7.B}) mean that
any scaling behaviour exhibited by the fluctuations
at equilibrium must also be exhibited by the actual
fluctuations for the non-equilibrium problem.
That is the direct result of local equilibrium in the asymptotic
leads, and of overall neutrality in the system made up of
conductor plus leads.

The proportionality of {\it all} thermally induced noise
to ambient temperature $T$ is inevitable in the degenerate limit.
This has implications for understanding shot noise in metallic
conductors
\cite{upon99,ithaca,gdcond}.
Shot noise never scales with $T$.
We comment further on the scaling issue in the later sections.

One may compare the analysis in terms of $G$ and $\Delta f^{(2)}$
with the equal-time correlator introduced by Gantsevich {\it et al.}
\cite{ggk}.
The equal-time correlator is itself a hierarchical
functional of other correlators (such as the current fluctuations),
whose solutions are unknown {\it a priori}
and which must be closed by force, either by truncation
or by an {\it ad hoc} heuristic device such as Boltzmann-Langevin.
This makes for a less-than-tractable computational scheme,
at least beyond a narrow repertoire of special limits
(linear response; classical particles; weak non-uniformity).
In particular, a calculable strategy for degenerate
non-equilibrium fluctuations, based on the equal-time correlator,
has yet to be demonstrated.

By contrast, we show below that $G$ is
explicitly determined by the dynamical Green function for the
linearized Boltzmann equation [see equation (\ref{cvlv5})].
As pointed out by Stanton
\cite{sw0},
the Boltzmann-Green functions are much more straighforward to
compute for a wide range of collision models. This ease of
calculation extends to $G$ and hence to $\Delta f^{(2)}$,
which provides the initial conditions
for a naturally closed dynamical solution.

Global charge neutrality requires that the total
fluctuation strength over the sample,
$\Delta N = {\sum}_{\bf r} \Omega({\bf r})
\langle \Delta f ({\bf r}) \rangle$, be conserved.
This constrains not only the steady-state but also the
time-dependent fluctuations.

\subsection{Time Dependence}

Calculation of the dynamic response requires
the time-dependent Green function
\cite{kogan}

\begin{equation}
R_{\alpha \alpha'}(t - t')
\buildrel \rm def \over =
\theta(t - t')
{ {\delta f_{\alpha}(t)}\over
  {\delta f_{\alpha'}(t')} },
\label{AX8}
\end{equation}

\noindent
with initial value
$R_{\alpha \alpha'}(0) = \delta_{\alpha \alpha'}$.
As with $G$, the variation is restricted.
The linearized Boltzmann equation satisfied by $R(t - t')$
is derived from equation (\ref{AX1})
and takes the form

\begin{equation}
\sum_{\beta}
{\left\{
\delta_{\alpha \beta} {{\partial}\over {\partial t}}
+ B[W^A f]_{\alpha \beta}
\right\}} R_{\beta \alpha'}(t - t') =
\delta(t - t') \delta_{\alpha \alpha'}.
\label{drdt}
\end{equation}

\noindent
Summation over $\alpha$ on both sides of this equation gives
zero contribution from
$\sum_{\alpha} \sum_{\beta}B[W^Af]_{\alpha \beta}R_{\beta \alpha'}(t-t')$.
Subsequent integration over $t$ leads to
conservation of normalization
\cite{kogan}:

\begin{equation}
\sum_{\alpha} R_{\alpha \alpha'}(t - t') = \theta(t - t').
\label{rgf1}
\end{equation}

\noindent
The time-dependent propagator is a two-point correlation.
It tracks the history of a fluctuation of unit strength,
created in state $\alpha'$ at time $t'$.
The fluctuation strength in state $\alpha$,
at the later time $t$, is
$R_{\alpha \alpha'}(t - t')$. In the long-time limit
equation (\ref{drdt}) goes to the steady-state equation (\ref{AX7.B})
independently of $\alpha'$, so that
$R_{\alpha \alpha'}(t \to \infty) \propto \Delta f_{\alpha}$.
Together with equation (\ref{rgf1}) this gives
\cite{kogan}

\begin{equation}
R_{\alpha \alpha'}(t \to \infty)
= { {\Delta f_{\alpha}}\over {\Delta N}}.
\label{rgf2}
\end{equation}

All of the time-dependent fluctuation properties
induced by the thermal background
are specified in terms of $R$
and the steady-state non-equilibrium fluctuation $\Delta f$.
From the dynamical particle-hole propagator
\cite{ggk},
that is $\Delta f_{\alpha \alpha'}^{(2)}(t) \equiv
R_{\alpha \alpha'}(t) \Delta f_{\alpha'}$,
one constructs the lowest-order moment

\begin{equation}
\Delta f_{\alpha}(t)
= \sum_{\alpha'} {\Delta f^{(2)}_{\alpha \alpha'}}(t)
\label{rgf3}
\end{equation}

\noindent
in analogy with equation (\ref{AX7}).
Equation (\ref{drdt}), with its adjoint
\cite{kogan},
implies that $\Delta f_{\alpha}(t) = \Delta f_{\alpha}$ for $t > 0$.
Thus the intrinsic time dependence of $\Delta f^{(2)}(t)$ is
not revealed through this quantity
\cite{fg1warn}.
Equation (\ref{rgf1}) ensures constancy of
the total fluctuation strength:
${\sum}_{\bf r} \Omega({\bf r})
\langle \Delta f ({\bf r}, t) \rangle = \Delta N$
for $t > 0$.

\subsection{Dynamic Correlations}

We move to the frequency domain. An important outcome is the
quantitative link between fluctuations
and resistive power dissipation in the non-equilibrium regime.
This requires expressing both the difference function $g$
and the adiabatic propagator $G$
directly in terms of the dynamical Green function.
The Fourier transform
${\rm R}(\omega) = \int dt e^{i\omega t} R(t)$
of the retarded time-dependent Green function satisfies

\begin{equation}
\sum_{\beta}
{\left\{ B[W^A f]_{\alpha \beta}
- i\omega \delta_{\alpha \beta} \right\}}
{\rm R}_{\beta \alpha'}(\omega) =
\delta_{\alpha \alpha'},
\label{AXX}
\end{equation}

\noindent
showing that ${\rm R}(\omega)$ is the resolvent for the
linearized Boltzmann operator of equation (\ref{bop}).
From equation (\ref{rgf1}), the global condition on
the resolvent is

\begin{equation}
\sum_{\alpha} {\rm R}_{\alpha \alpha'}(\omega)
= -{1\over {i(\omega + i\eta)} } {~~~~~~ }(\eta \to 0^+).
\label{rgf5}
\end{equation}

\noindent
At face value this does not match the corresponding
criterion for $G$, equation (\ref{dgauss}).
To solve equation (\ref{AXG})
for the steady-state propagator explicitly
in terms of the dynamic one, we follow Kogan and Shul'man
\cite{kogan}
in introducing the intrinsically correlated part of
${\rm R}(\omega)$. This is

\begin{equation}
{\rm C}_{\alpha \alpha'}(\omega) = {\rm R}_{\alpha \alpha'}(\omega)
+ {1\over {i(\omega + i\eta)} }
{ {{\Delta f}_{\alpha}}\over {\Delta N} }.
\label{f2.2}
\end{equation}

\noindent
Once the long-time adiabatic term is removed,
${\rm C}(\omega)$ conveys the purely transient response
of the system.
It satisfies a pair of identities
\cite{kogan}.
First, the Fourier transform of the relation
$\Delta f(t) = \theta(t)\Delta f$ translates to

%\begin{mathletters}
%\label{cckt}
\begin{equation}
\sum_{\alpha'} {\rm C}_{\alpha \alpha'}(\omega)
\Delta f_{\alpha'} = 0 {~~} 
{\rm for {~}all} {~}\alpha,
\label{cckta}
\end{equation}

\noindent
while equation (\ref{rgf5}) leads to

\begin{equation}
\sum_{\alpha} {\rm C}_{\alpha \alpha'}(\omega) = 0 {~~} 
{\rm for {~}all} {~}\alpha'.
\label{ccktb}
\end{equation}
%\end{mathletters}

\noindent
The latter parallels the constraint on $G$.
Like ${\rm R}(\omega)$, the correlated propagator
is analytic in the upper half-plane ${\rm Im}\{\omega\} > 0$,
and satisfies the Kramers-Kr\"onig dispersion relations.
Unlike ${\rm R}(\omega)$, however, ${\rm C}(\omega)$
is regular for $\omega \to 0$.

We now obtain $g$ and $G$ in terms of the
correlated dynamical response. Consider the equation

\begin{equation}
%IOP+ \fl
\sum_{\alpha'}
  {\left\{
B[W^A f]_{\alpha \alpha'} - i\omega \delta_{\alpha \alpha'}
  \right\}}
{\rm g}_{\alpha'}(\omega)
= { { e{\bf {\widetilde E}}({\bf r}) }\over {\hbar} }{\bbox \cdot}
  { {\partial f^{\rm eq}_{\alpha}}\over {\partial {\bf k}} }
+ \sum_{\alpha'}
{g}_{\alpha} W^A_{\alpha \alpha'}
{g}_{\alpha'};
\label{cvlv1}
\end{equation}

\noindent
inversion with the resolvent yields

\begin{equation}
{\rm g}_{\alpha}(\omega)
= \sum_{\alpha'}
{\rm C}_{\alpha \alpha'}(\omega)
{ e{\bf {\widetilde E}} ({\bf r'})\over \hbar } {\bbox \cdot}
{ {\partial f^{\rm eq}_{\alpha'}}\over {\partial {\bf k'}} }
+ \sum_{\alpha' \beta}
{\rm C}_{\alpha \alpha'}(\omega)
{g}_{\alpha'} W^A_{\alpha' \beta} {g}_{\beta}.
\label{AX15.3}
\end{equation}

\noindent
The dominant low-frequency component of ${\rm R}(\omega)$
does not contribute to the right-hand side of this equation.
In the first term it results in a decoupling of
the summation over ${\alpha'}$, yielding zero
because $\partial f^{\rm eq}_{\alpha'}/\partial {\bf k'}$
is odd in ${\bf k'}$; in the second term, decoupling means that
the double summation over ${\alpha'}$ and ${\beta}$
vanishes by antisymmetry.
In the static limit equation (\ref{cvlv1}) becomes
the inhomogeneous equation (\ref{AX5}); moreover
equation (\ref{ccktb}) means that ${\rm g}(\omega = 0)$
satisfies equation (\ref{gauss}),
the sum rule for $g$. Therefore $g = {\rm g}(0)$, or

\begin{equation}
g_{\alpha}
= \sum_{\alpha'}
{\rm C}_{\alpha \alpha'}(0)
{ e{\bf {\widetilde E}} ({\bf r'})\over \hbar } {\bbox \cdot}
{ {\partial f^{\rm eq}_{\alpha'}}\over {\partial {\bf k'}} }
+ \sum_{\alpha' \beta}
{\rm C}_{\alpha \alpha'}(0) g_{\alpha'} W^A_{\alpha' \beta} g_{\beta}.
\label{cvlv2}
\end{equation}

\noindent
This identity is central to the fluctuation-dissipation theorem.

In models with symmetric scattering,
$W^A$ is zero and the adiabatic Green function assumes
a simple form on varying both sides of equation (\ref{cvlv2}):

\begin{equation}
G_{\alpha \alpha'} = {\rm C}_{\alpha \alpha'}(0)
{ e{\bf {\widetilde E}} ({\bf r'})\over \hbar } {\bbox \cdot}
{ {\partial}\over {\partial {\bf k'}} }.
\label{AXG2}
\end{equation}

\noindent
More generally, an analysis similar to that for
${\rm g}(\omega)$ can be used directly
for the adiabatic propagator.
Introduce the operator ${\rm G}(\omega)$, defined to
satisfy the dynamic extension of equation (\ref{AXG}),

\begin{equation}
%IOP+ \fl
\sum_{\beta}
  {\left\{
B[W^A f]_{\alpha \beta} - i\omega \delta_{\alpha \beta} 
  \right\}}
{\rm G}_{\beta \alpha'}(\omega)
=
\delta_{\alpha \alpha'}
\!{\left[
{ { e{\bf {\widetilde E}} ({\bf r'})}
\over {\hbar} }{\bbox \cdot}
   { {\partial }\over {\partial {\bf k'}} }
+ \sum_{\beta} W^A_{\beta \alpha'}
{g}_{\beta}
\right]}\!- W^A_{\alpha \alpha'}
{g}_{\alpha}.
\label{AYG}
\end{equation}

\noindent
This has the solution

\begin{equation}
{\rm G}_{\alpha \alpha'}(\omega)
= {\rm C}_{\alpha \alpha'}(\omega)
{ { e{\bf {\widetilde E}} ({\bf r'})}
\over {\hbar} }{\bbox \cdot}
   { {\partial }\over {\partial {\bf k'}} }
- \sum_{\beta}
{\Big(
{\rm C}_{\alpha \alpha'}(\omega)
-
{\rm C}_{\alpha \beta}(\omega)
\Big)}
W^A_{\alpha' \beta} {g}_{\beta}.
\label{cvlv4}
\end{equation}

\noindent
In the first term on the right-hand side, the low-frequency
component of ${\rm R}(\omega)$ makes no contribution after decoupling
because the physical distributions $F_{\alpha}$
on which ${\rm G}(\omega)$ operates vanish sufficiently fast that
${\sum}_{\bf k} {\partial F}_{\alpha}/{\partial {\bf k}} = {\bf 0}$.
In the second right-hand term the uncorrelated parts of
${\rm R}_{\alpha \alpha'}(\omega)$
and ${\rm R}_{\alpha \beta}(\omega)$
cancel directly. We conclude as before that

\begin{equation}
G_{\alpha \alpha'}
= {\rm C}_{\alpha \alpha'}(0)
{ { e{\bf {\widetilde E}} ({\bf r'})}
\over {\hbar} }{\bbox \cdot}
   { {\partial }\over {\partial {\bf k'}} }
- \sum_{\beta}
{\Big(
{\rm C}_{\alpha \alpha'}(0)
-
{\rm C}_{\alpha \beta}(0)
\Big)}
W^A_{\alpha' \beta} g_{\beta}.
\label{cvlv5}
\end{equation}

\noindent
This is a crucial result. It shows
(i) that the adiabatic structure of the steady state, through $G$,
is of one piece with the correlated dynamic response
(the result of causality and global charge neutrality), and
(ii) that the non-equilibrium correlation
structure evolves {\it expressly}
out of the equilibrium state, through the
specific functional form of $G \Delta f^{\rm eq}$.

We have proved that this non-perturbative kinetic
description of fluctuations is self-contained,
given its conventional set of assumptions
and boundary conditions.
The kinetic formalism has inherent predictive power.
Hence, extraneous phenomenologies are not needed
to make it viable. This is in sharp distinction
to the diffusive Boltzmann-Langevin viewpoint
\cite{kogancpu}.

\subsection{Spectral Density}

The vehicle for the physics of current noise is
the velocity auto-correlation.
It is a two-point distribution in real space, built on
the correlated part of the two-particle fluctuation
$\Delta {\rm f}^{(2)}_{\alpha \alpha'}(\omega)
= {\rm R}_{\alpha \alpha'}(\omega) \Delta f_{\alpha'}$.
Following Gantsevich, Gurevich, and Katilius
\cite{ggk}
it is a double sum over the kinematic states:

\begin{equation}
%IOP+ \fl
{ \langle\!\langle {\bf v} {\bf v'}
{\Delta {\rm f}^{(2)} } ({\bf r}, {\bf r'}; \omega) 
\rangle\!\rangle}_{\rm c}'
\buildrel \rm def \over =
{1\over \Omega({\bf r})}
\sum_{{\bf k}, s}
{1\over \Omega({\bf r'})}
\sum_{{\bf k'}, s'}
{\bf v}_{{\bf k} s}
 {{\rm Re} \{{\rm C}_{\alpha \alpha'}(\omega)\} } 
{\bf v}_{{\bf k'} s'}
\Delta f_{\alpha'}.
\label{AXY}
\end{equation}

\noindent
Its physical meaning is the following. At
any time, the system in steady state has a fluctuation 
background that is fed by spontaneous energy exchanges
with the (equilibrium) thermal bath.
The average strength of the fluctuations
is fixed by the distribution $\Delta f$.
The elementary modes making up this background
are long-wavelength electron-hole excitations; these are
given by $\Delta {\rm f}^{(2)}(\omega)$. The pair excitations
are not themselves dynamically stable.
Their transient evolution is determined by the
propagator ${\rm C}(\omega)$ acting upon the
ensemble-averaged background source, $\Delta f$.
Finally, the velocity-velocity correlation
for the pair process is obtained
by attaching velocity operators at the start and end
of the electron-hole excitations,
and summing over states
\cite{pinoz}.

This approach to auto-correlations makes straightforward,
and completely standard, use of the Boltzmann
\cite{ggk}
and Fermi-liquid
\cite{pinoz}
theories.
Our particular contribution is to have given an explicit recipe
for computing the steady-state form of $\Delta f$ semi-classically,
by analyzing the underlying adiabatic propagator $G$.
Practical calculations should thereby become easier
for degenerate systems at high driving fields.

The one-point object derived from equation (\ref{AXY}),

\begin{equation}
S_f ({\bf r}, \omega)
= e^2 \sum_{\bf r'} \Omega({\bf r'})
{\langle\!\langle 
( {\bf {\widetilde E}}({\bf r}){\bbox \cdot}{\bf v} )
( {\bf {\widetilde E}}({\bf r'}){\bbox \cdot}{\bf v'} )
{\Delta {\rm f}^{(2)}({\bf r}, {\bf r'}; \omega) }
\rangle\!\rangle}_{\rm c}',
\label{Svv}
\end{equation}

\noindent
measures the local effect of fluctuations that are spread
throughout the system. Formally it is the auto-correlation
function of the power transferred from field to carriers,
an inherently volume-distributed property that is represented
here in terms of a locally defined spectral density.
$S_f$ is closely related to the thermally induced current
noise, integrated over the entire structure.
In the weak-field limit it satisfies the FDT.

The two-point velocity correlator

\[
{ \langle\!\langle {\bf v} {\bf v'}
{\Delta {\rm f}^{(2)} } ({\bf r}, {\bf r'}; \omega) 
\rangle\!\rangle}_{\rm c}'
/\Delta N,
\]

\noindent
which is the response to a unit change of total particle number
(and which does not scale with $T$), 
should provide the direct basis for shot-noise calculations
across distances $|{\bf r} - {\bf r'}|$
comparable to the mean free path.
It is natural to ask how shot noise fits into the framework of
equation (\ref{Svv}). Within semi-classical kinetics, the short
answer is that shot noise {\it cannot} be
encompassed by the generic spectrum for thermal noise.
For, as we have rigorously shown, all thermal fluctuations
are required to scale with $T$ in the
strongly degenerate (metallic) regime.
Shot noise, on the other hand, has no such scaling.
Therefore, whatever the kinetic description
of shot noise may be, it is impossible for it to exhibit
the smooth physical ``cross-over'' into thermal noise
that is the primary feature of every diffusive model
\cite{kogancpu}.

Our approach to the kinetics of mesoscopic
shot noise is explored in reference
\onlinecite{gdcond}.
In reference
\onlinecite{upon99}
we propose a quite specific experimental test of our theory.
The new predictions made there are in stark
contradistinction to the diffusive ones.

\subsection{Fluctuation and Dissipation}

The fluctuation-dissipation relation near equilibrium ties
the spectral density of the thermal current fluctuations to the
dissipative effects of the steady current in the system.
However, dissipation by itself does not exhaust the physics of
this sum rule. There are non-linear terms,
negligible in linear response, that
dominate the high-field behaviour of the noise
\cite{sw2,gc}.
In view of this, it is imperative to
reveal the precise nature and action of these terms.
We do so.

The resolvent property of ${\rm R}(\omega)$
provides a formal link between the
steady-state (one-body) solution $g$ and
the dynamical (two-body) fluctuation
$\Delta {\rm f}^{(2)}$ at the semi-classical level.
Taken to its equilibrium limit this becomes the familiar theorem.
The connection is made in two steps.
Consider the kinematic identity

\begin{equation}
{ {\partial f^{\rm eq}_{\alpha}}\over {\partial {\bf k}} }
= -{ \hbar \over {k_{\rm B}T} }
{\bf v}_{{\bf k} s}
\Delta f^{\rm eq}_{\alpha}
\label{AX14}
\end{equation}

\noindent
and apply it to the leading term
on the right-hand side of equation (\ref{cvlv2}). The result is

\begin{equation}
g_{\alpha} = -{ {e}\over {k_{\rm B} T} }
\sum_{\alpha'}
{\rm C}_{\alpha \alpha'}(0)
( {\bf {\widetilde E}} {\bbox \cdot} {\bf v} )_{\alpha'}
\Delta f^{\rm eq}_{\alpha'} + h_{\alpha},
\label{AX15}
\end{equation}

\noindent
in which $h_{\alpha} = {\sum}_{\alpha' \beta}
{\rm C}_{\alpha \alpha'}(0) g_{\alpha'} W^A_{\alpha' \beta} g_{\beta}$.
Evaluation of the current density according to
${\bf J}({\bf r}) = -e\langle {\bf v} g \rangle$,
means that the power density
$P({\bf r}) = {\bf {\widetilde E}} ({\bf r}) {\bbox \cdot} 
{\bf J}({\bf r})$ for Joule heating can be written as

\begin{equation}
P({\bf r}) =
{ {e^2}\over {k_{\rm B} T} }
{1\over \Omega({\bf r})}
\sum_{{\bf k},s}
\sum_{\alpha'}
( {\bf {\widetilde E}} {\bbox \cdot} {\bf v} )_{\alpha}
{\rm C}_{\alpha \alpha'}(0)
( {\bf {\widetilde E}} {\bbox \cdot} {\bf v} )_{\alpha'}
\Delta f^{\rm eq}_{\alpha'}
- e{\langle {\bf {\widetilde E}} {\bbox \cdot} {\bf v} h \rangle}.
\label{AY16}
\end{equation}

In the second step we take the one-point spectral
function $S_f$ in the static limit,
substituting for $\Delta f$ from equation (\ref{AX7})
in the right-hand side of equation (\ref{Svv}) to give

\begin{equation}
S_f ({\bf r}, 0) =
{e^2\over \Omega({\bf r})}
\sum_{{\bf k},s}
\sum_{\bf r'} \sum_{{\bf k'},s'}
( {\bf {\widetilde E}} {\bbox \cdot} {\bf v} )_{\alpha}
{\rm C}_{\alpha \alpha'}(0)
( {\bf {\widetilde E}} {\bbox \cdot} {\bf v} )_{\alpha'}
\Delta f^{\rm eq}_{\alpha'}
+ S_g ({\bf r}, 0),
\label{eq17}
\end{equation}

\noindent
where $S_g({\bf r}, 0)$ is generated by replacing
$\Delta f$ with $\Delta g \equiv \Delta f - \Delta f^{\rm eq}$
in equation (\ref{AXY}), and subsequently in equation (\ref{Svv}).
Direct comparison of equations (\ref{AY16}) and (\ref{eq17}) leads to

\begin{equation}
{ {S_f ({\bf r}, 0)}\over {k_{\rm B} T} }
= P({\bf r})
+ e{\langle {\bf {\widetilde E}} {\bbox \cdot} {\bf v} h \rangle}
+ { {S_g ({\bf r}, 0)}\over {k_{\rm B} T} }.
\label{eq18}
\end{equation}

\noindent
This is the precise connection between the non-equilibrium
thermal current fluctuations and resistive dissipation in the system.

The limiting weak-field form of equation (\ref{eq18}) is easily obtained.
We prove that it is the linear fluctuation-dissipation theorem.
Observe that the term in $h$ on the right-hand side
varies as ${\widetilde E} g^2$, while the final term varies as
${\widetilde E}^2 \Delta g$;
both of these contributions are therefore of order
${\widetilde E}^3$.
Suppose that the system is uniform. Then
${\bf {\widetilde E}} = {\bf E}_{\rm ext} = {\bf E}$
acts along the {\it x}-axis.
Division by $E^2$ on both sides of equation (\ref{eq18}) gives

\begin{equation}
{1\over E^2}
{ {S_f ({\bf r}, 0)}\over k_{\rm B}T}
{~\rightarrow~} {|J_x|\over E}
= \sigma,
\label{eq19}
\end{equation}

\noindent
where $\sigma$ is the low-field conductivity.
Equation (\ref{eq19}) is the canonical FDT.

The non-dissipative and purely non-equilibrium structures
beyond $P({\bf r})$ can be expanded similarly to it.
We discuss the symmetric-scattering case,
for which there is no contribution
$e\langle {\bf {\widetilde E}} {\bbox \cdot} {\bf v} h \rangle$.
Within $S_g$ we apply the formula for the adiabatic 
propagator, equation (\ref{AXG2}), to express
$\Delta g = \sum G \Delta f^{\rm eq}$
in terms of the correlated dynamic response function
${\rm C}(\omega)$. This produces the closed form

\begin{eqnarray}
%IOP+ \fl
S_g ({\bf r}, 0)
%IOP+ &=
=&&
{e^2\over \Omega({\bf r})} \sum_{{\bf k}, s} 
\sum_{\beta}
( {\bf {\widetilde E}} {\bbox \cdot} {\bf v} )_{\alpha}
{\rm C}_{\alpha \beta}(0)
( {\bf {\widetilde E}} {\bbox \cdot} {\bf v} )_{\beta}
{\left(
\sum_{\alpha'}
{\rm C}_{\beta \alpha'}(0)
{ {e {\bf {\widetilde E}}} ({\bf r'}) \over \hbar} {\bbox \cdot}
{ {\partial \Delta f^{\rm eq}_{\alpha'}}\over
{\partial {\bf k'}} }
\right)}
\cr
{\left. \right.} \cr
%IOP+ &=
=&&
-{e^3\over {k_{\rm B}T}} {1\over \Omega({\bf r})}
\sum_{{\bf k}, s}
\sum_{\alpha'}
( {\bf {\widetilde E}} {\bbox \cdot} {\bf v} )_{\alpha}
   {
(
{\rm C}(0)
{\bf {\widetilde E}} {\bbox \cdot} {\bf v}
)
}^2_{\alpha \alpha'}
(1 - 2f^{\rm eq}_{\alpha'}) \Delta f^{\rm eq}_{\alpha'}.
\label{eq22b}
\end{eqnarray}

\noindent
The second line follows from the first after
using equation (\ref{AX14}) to express
${\partial \Delta f^{\rm eq}}/{\partial {\bf k}}$
in terms of $f^{\rm eq}$ and $\Delta f^{\rm eq}$,
and taking an inner sum into
$( {\rm C}(0) {\bf {\widetilde E}} {\bbox \cdot} {\bf v} )^2$.

The expression above differs markedly from
the rate of energy loss $P({\bf r})$ by Joule heating.
In contrast, $S_g ({\bf r}, 0)$ relates directly to
non-equilibrium broadening of the fluctuations, due to
the excess energy gained from the field
during intervals of ballistic flight
\cite{sw0}.
The {\it extent} of the broadening is limited by
dynamical dissipation of the excess energy,
locally (by prompt inelastic scattering) or remotely
(by carrier relaxation in the ideally absorbing terminals).
The impact of this term on current noise is felt only for
substantial departures from the weak-field regime.

There exist several alternative generalizations of the FDT
for extended bulk systems
\cite{vvt,nougier,nerlul}.
We mention the best known,
which defines the non-equilibrium noise temperature $T_{\rm n}$
pivotal to the interpretation of device-noise data
\cite{nougier}.
Phenomenologically $T_{\rm n}$ is obtained,
for a non-linear operating point, by normalizing $S_f$ with
the  differential conductivity $\sigma_x({\widetilde E})
= \partial J_x / \partial {\widetilde E}_x$ such that
$k_{\rm B} T_{\rm n}({\widetilde E}) \equiv
S_f/\sigma_x({\widetilde E}) {\widetilde E}_x^2$,
corresponding to the output of a small-signal noise measurement.
(In general $T_{\rm n}$ is not isotropic.)
Our equations (\ref{AY16}) -- (\ref{eq22b})
provide a microscopic framework for computing
the noise spectral density
in a wide class of degenerate systems.
Since $\sigma_x({\widetilde E})$ is also
calculable within the same framework, this yields $T_{\rm n}$.

\section{Application to High-Field Noise}

We can now explore one of the most significant properties
of the excess spectrum $S_g$: its strong inhibition by degeneracy.
That there exists an additional, purely quantum-statistical,
constraint on field-driven broadening
is seen directly in the factor $(1 - 2f^{\rm eq})$
of equation (\ref{eq22b}). This suppresses the
contribution of $S_g$ relative to the
corresponding classical result, in which the
factor is unity. Suppression of electron heating
by Pauli exclusion reflects the large energy cost of displacing
electrons deep inside the Fermi sea.

To highlight the difference between dissipative and hot-electron
terms, we revisit a simple example
\cite{gc,mbix},
the uniform electron gas in the
constant-collision-time (Drude) approximation
subject to a field ${\bf E} = -E{\bf {\hat x}}$.
Expressions for the power density $P$
and hot-electron component $S_g$ are derived in the Appendix.
The thermally driven current-current spectral density, taken
over a uniform sample of length $L_x$ and total volume $\Omega$,
is given by
\cite{nougier}

\begin{eqnarray}
{\cal S}(E, \omega)
%IOP+ \buildrel \rm def \over &=
\buildrel \rm def \over =&&
4 \sum_{\bf r}\Omega({\bf r}) \sum_{\bf r'}\Omega({\bf r'})
{\left\langle\!\!\left\langle
\left( -{ev_x\over L_x} \right) \left( -{ev'_x\over L_x} \right)
{\Delta {\rm f}^{(2)} } (\omega) 
\right\rangle\!\!\right\rangle}_{\rm c}'
\cr
{\left. \right.} \cr
%IOP+ &=
=&&
4 { {\Omega S_f(\omega)}\over {L_x^2 E^2} }.
\label{drude0}
\end{eqnarray}

\noindent
Writing the sample conductance as ${\cal G} = \Omega P/L_x^2 E^2$,
the static limit of the spectrum is determined by equation (\ref{eq18}):

\begin{equation}
{\cal S}(E,0) = 4{\cal G}k_{\rm B}T
{\left[ 1 + { {S_g(0)}\over{Pk_{\rm B}T} } \right]}
= 4{\cal G}k_{\rm B}T
{\left[ 1 + {{\Delta {n}}\over {n}}
{\left( { {m^* \mu_{\rm e}^2 E^2}\over k_{\rm B}T } \right)} \right]}.
\label{drude1}
\end{equation}

\noindent
We have substituted for $P$ and $S_g$ respectively from
equations (\ref{apxT8}) and (\ref{apxT10}).
The electronic density is ${n}$ while
$\Delta {n} = \Delta N/\Omega$ is the number-fluctuation density.
The effective electron mass is $m^*$ and $\mu_{\rm e}$ is the mobility.

The term $S_g/Pk_{\rm B}T$ is a relative measure of the hot-electron
contribution to the noise.
The inhibiting effect of degeneracy, through ${\Delta {n}}/{n}$,
is greatest at low temperature and least in the classical regime.
When the Fermi energy $\varepsilon_{\rm F}$ satisfies
$\varepsilon_{\rm F} \ll k_{\rm B}T$,
the ratio $\Delta n/n$ goes to unity and
the hot-electron term is that of a classical
electron gas (low density, high temperature). Its
form in the high-field limit $E \gg \sqrt{k_{\rm B}T/m^* \mu_{\rm e}^2}$
is ${\cal S} \sim 4{\cal G}m^*\mu_{\rm e}^2E^2$,
asymptotically independent of $T$.

On the other hand, when
$k_{\rm B}T \ll \varepsilon_{\rm F}$ the system
is strongly degenerate.
In a $\nu$-dimensional system
we have $\varepsilon_{\rm F} \propto n^{2/\nu}$. Then

%\begin{equation}
\[
{{\Delta {n}}\over {n}} = {k_{\rm B}T\over {n}}
{{\partial {n}}\over {\partial \varepsilon_{\rm F}}}
\to {{\nu k_{\rm B}T}\over {2 \varepsilon_{\rm F}}};
\]
%\label{drude2}
%\end{equation}

\noindent
with equation (\ref{drude1}) this leads to

\begin{equation}
{ {{\cal S}(E,0)}\over {{\cal S}(0,0)}}
\to 1 + {\nu\over 2}
{\left( { {m^* \mu_{\rm e}^2 E^2}\over \varepsilon_{\rm F} } \right)}.
\label{drude5}
\end{equation}

\noindent
Note that the thermal fluctuation spectrum
${\cal S}(E,0)$ {\it necessarily}
vanishes with temperature, while its ratio
with the Johnson-Nyquist
spectral density ${\cal S}(0,0) = 4{\cal G}k_{\rm B}T$
continues to exhibit a hot-electron excess which is
now scaled by the Fermi energy.

Figure 1 displays the excess-noise
spectral ratio in a two-dimensional electron gas,
as a function of the applied field,
when $T$ ranges from the degenerate limit to
well above the Fermi temperature
$T_{\rm F} = \varepsilon_{\rm F}/k_{\rm B}$.
For $T$ much greater than both $T_{\rm F}$ and
$m^*\mu_{\rm e}^2E^2/k_{\rm B}$
the excess contribution becomes classical,
independent of temperature, and thus small
compared with the now-dominant base value
${\cal S}(0,0)$. This is evident in figure 1
through the gradual downward shift of
the plots, with increasing $T$.

Equation (\ref{drude5}) may be compared with a
perturbative estimate by Landauer
\cite{landauer}
in the degenerate limit,
for which the analogous excess term is
$(\delta U/ k_{\rm B}T)^2$, where
$\delta U \sim m^* \mu_{\rm e} E v_{\rm F}$
is a characteristic energy gain.
Taken at face value, this would suggest that hot-electron effects
in the low-$T$ regime can be enhanced even more by
further cooling of the system.

This counter-intuitive result comes from inappropriate
use of perturbation analysis.
Series expansion of the thermal current noise,
in powers of $E$, fails to account for
non-analyticity of the full non-perturbative solution
in its approach to equilibrium
\cite{bakshi}.
Non-analyticity of the distribution function $f_{\bf k}$ 
precludes the reliable calculation of moment averages
by expanding about equilibrium, as in reference \cite{landauer}.
(It is reassuring -- and only seemingly fortuitous -- that the
actual linear current response is reproduced
exactly by solving
the transport equation, as usual, to first order in the field
\cite{sw0}.)

The relevance of non-analyticity to transport physics
has been questioned by Kubo, Toda, and Hashitsume
\cite{kubo}.
They regard its appearance as spurious,
a specific artefact of the crude way in which
the Drude approximation treats real collisions.
That is to overlook the appreciably broader
evidence for non-analyticity in the variety
of collision models assembled by Bakshi and Gross
\cite{bakshi}.

Even within the Drude model of a degenerate conductor
(over-simplified though it is), it is clear that its
exact non-perturbative solution does
produce physically consistent scaling of
the excess noise with $T$.
Equally clearly, finite-order response theory does not.
Kubo linear response recovers only ${\cal S}(0,0)$
and misses the non-linear excess noise altogether.
Such sharp differences
between perturbative and non-perturbative predictions
should be experimentally measurable in the hot-electron
spectrum. In our view, issues of non-analyticity
and its physical manifestation remain open.

We make some final comments on shot noise and the impossibility
\cite{upon99,ithaca,gdcond}
of a {\it theoretical} cross-over, unifying thermal
and shot noise for mesoscopic metallic wires.
The diffusive cross-over formula, ostensibly identical in form
to equation (\ref{drude0}),
{\it always} generates a $T$-independent term
\cite{but,beebut,thmldr,nagaev1,djb2}.
One might have expected that a computation of
the spectral density of equation (\ref{drude0}),
taken in the semi-classical quasi-ballistic limit
$L_x \ll m^*v_{\rm F}\mu_{\rm e}/e$,
would yield an expression for ${\cal S}(E,0)$ that is
independent of $T$ and proportional to the current $I = {\cal G}V$;
in other words, shot noise
\cite{sw0}.

We have carried out this quasi-ballistic exercise
for a degenerate system, in simplified form
\cite{mbix}.
At high fields it gives
${\cal S}(E,0) \sim 2eI(k_{\rm B}T/\varepsilon_{\rm F})$.
This is indeed linear in $I$ but thermal nevertheless,
since its immediate source is
the generic spectral relation, equation (\ref{eq22b}).

Thermal fluctuations are induced by spontaneous and
quasi-continuous changes in the total internal energy of
carriers, throughout the whole active volume of a device.
Shot-noise fluctuations are induced by spontaneous and
{\it discrete} changes in total carrier number, through the
device's {\it interfaces} with the outer circuit.
Such qualitative and topological distinctions may be
of little practical importance in the classical macroscopic world.
However, it is not at all clear that they are
immaterial to the metallic mesoscopic regime.
The issue is under active examination
\cite{gdcond}.

A leading task is to identify the kinetic origin of
the {\it empirical} cross-over between thermal and shot noise,
apparent in real mesoscopic conductors
\cite{liefr,rez,ksgje,smd,sbkpr}.
Once again we stress that, regardless of how shot noise
is to be described microscopically, the logical and conceptual
gaps between diffusive explanations of the
cross-over (quantum as well as semi-classical)
\cite{theor,blbu,kogancpu}
and strictly conventional kinetic theory
have already been uncovered, characterized and analyzed
\cite{upon99,ithaca}.
We will present a fully detailed semi-classical
kinetic model of shot noise in due course.

\section{Summary}

We have described, and applied, a genuinely non-equilibrium
kinetic formalism for current fluctuations.
It holds for metallic systems down to mesoscopic scales,
within the ambit of semi-classical theory.
Our strategy for incorporating microscopic Fermi-liquid
correlations within the Boltzmann picture
safeguards the conservation laws at the two-body level.
Conservation continues to underpin the nature of
current noise at high fields.

Our theory leads to a precise quantitative link between
non-equilibrium thermal current fluctuations and energy dissipation.
In its low-field form, this is the standard
fluctuation-dissipation theorem of linear-response analysis.
At high fields, it highlights the pervasiveness of strong degeneracy
even in hot-electron noise.

In a completely standard description, such as ours,
correctness and calculability do not issue from the {\it ad hoc}
assumption of fictive Langevin noise sources;
even less do they rely on diffusive analogies that fail
manifestly to respect
the canonical sum rules in electronic Fermi systems.
Rather, the model's integrity will stem from the microscopic
structure of its underlying Green functions. They describe
how the fundamental electron-hole pair excitations evolve
within a metallic conductor, in the semi-classical limit.

We have discussed how to map these native polarized correlations
non-perturbatively, from their equilibrium distribution
to its analogue in the externally driven conductor.
The resulting high-field noise spectrum yields a faithful
signature of its source: the elementary non-equilibrium
electron-hole polarization processes.

The main, and physically inevitable, consequence of this
{\it rigidly orthodox} kinetic investigation is the
intrinsic scaling of degenerate-electron fluctuations
with thermodynamic temperature. For diffusive phenomenologies,
this is one phenomenon too many. To sustain their
predictions for the shot noise of metallic wires,
they have no choice but to deny outright all possibility
of $T$-scaling for hot-electron noise
\cite{ithaca}.
This must be so for any theory that predicts
a seamless cross-over between thermal and shot noise.

There is an undeniable connection between $T$-scaling
and the dominance of long-range screening, as of degeneracy,
in the polarizable electron gas. Logical examination shows
that this nexus can be broken only by contradicting the
standard picture of charge fluctuations in metals.
Diffusively inspired models would seem to do exactly  that
\cite{csr,upon99,ithaca}.
So far, no such model has rationalized the heroic
departure from principles that have been understood,
widely and thoroughly, for some time
\cite{pinoz}.

We envisage two extensions to this work:
the systematic inclusion of {\it Coulomb screening} within the
microscopic structure of the fluctuations
\cite{gdii},
and the analysis of shot noise as a {\it kinetic process}
quite separate from thermally driven noise
\cite{gdcond}.
Coulomb effects are particularly evident
in strongly confined electron systems, such as the
two-dimensional electron gas in a III-V heterojunction quantum well
\cite{ferry}.
Self-consistent Coulomb screening in a confined
channel should markedly reduce the scale of
thermal fluctuations in the current.

Shot noise and thermal noise have disparate properties,
which no-one disputes.
Shot noise never scales with ambient temperature,
while excess thermal current noise must do so
if there is strong degeneracy. Coulomb effects too
may differentiate between the two kinds of fluctuations.
If so, then selective action of the Coulomb correlations
could serve as an experimental tool to distinguish between
excess thermal current noise and shot noise.
This would help to pin-point both the distinct sources
of non-equilibrium mesoscopic fluctuations and the disposition of
Coulomb forces at small scales.
We take up these themes in the forthcoming papers.

Every formalism for mesoscopic noise stands or falls
by its new predictions. Ours is no exception
\cite{upon99,gdii}.
Boltzmannian kinetics are obviously not equipped
to give the final word on quantum fluctuation effects;
be that as it may, it hardly needs saying that
{\it any} mesoscopic model, whatever its origin,
should be totally consistent with
the established physical facts.
In the context of the metallic electron gas, noise
descriptions which claim to be truly microscopic
must address full sum-rule consistency as a matter of course.
This transcends semi-classical analysis
and is by far our most important message.

\section*{Acknowledgments}

We are indebted to the late R Landauer for his generous
encouragement in the preliminary stages of this work.
We thank Erika Davies for help with calculations and figures
and N W Ashcroft, R J-M Grognard, K I Golden and
D Neilson for fruitful discussions.

\appendix

\section{Uniform Drude model}

We derive the dynamical fluctuation structure
for a single parabolic conduction band
with uniform electron density ${n}$ and
constant mobility $\mu_{\rm e} = e \tau / m^*$, where $\tau$ is the
spin-independent collision time and $m^*$ the effective mass.
The system is driven by a uniform field 
${\bf {\widetilde E}} = {\bf E} = -E {\bf {\hat x}}$
acting in the negative (drain to source) direction.
We take variations which are homogeneous over the sample region,
so that the fluctuations of interest have no spatial dependence.

The Boltzmann equation in the model is

\begin{equation}
\left[ { {\partial }\over {\partial t} } 
+  {
   { { eE}\over {\hbar} }
  }
   { {\partial }\over {\partial k_x} }
+ { 1\over \tau} \right] f_{\bf k}(t)
= 
{ {\langle f(t) \rangle} \over {\langle f^{\rm eq} \rangle} }
{ f_k^{\rm eq}\over \tau}.
\label{apxT1}
\end{equation}

\noindent
Since the Boltzmann operator is linear, the
fluctuation structure is qualitatively similar to that for
elastic scattering [differences
arise from the inhomogeneous term in $f^{\rm eq}$,
notably in the behaviours of $R(t)$ and $\Delta f(t)$].
We solve equation (\ref{apxT1})
by Fourier transforms in reciprocal space, so that the transform
$F_{\bbox{\rho}} \equiv \Omega^{-1}{\sum}_{\bf k} f_{\bf k}
\exp (i{\bf k}{\cdot}{\bbox{\rho}})$
of the steady-state distribution takes the form

\begin{equation}
F_{\bbox{\rho}} = { { F_{\bf 0} }\over { F^{\rm eq}_0 } }
 { { F^{\rm eq}_{\rho} } \over
 { 1 - ik_d \rho_x } },
\label{apxT2}
\end{equation}

\noindent
where $k_d = e E\tau/\hbar $
and $F_{\bf 0} = {1\over 2} \langle f \rangle$ per spin state.
While a formal distinction is made between $F_{\bf 0}$ and
$F^{\rm eq}_0$, the physical normalization is always
$F_{\bf 0} = F^{\rm eq}_0 = {1\over 2}{n}$.
Note also that $F_{\bbox{\rho}}$ is singular for $\rho_x = -ik_d^{-1}$.
In wave-vector space this means that $f_{\bf k}$ is non-analytic
at $E = 0$. The same can be said for $\Delta f_{\bf k}$. 

The transform of the dynamic response function,

\[
{\sf {\cal R}}_{\bbox{\rho} \bbox{\rho}' }(\omega) 
\equiv {1\over \Omega^2} {\sum}_{\bf k} {\sum}_{\bf k'}
{\rm R}_{{\bf k} {\bf k'}}(\omega)
\exp [i({\bf k}{\cdot}{\bbox{\rho}}
- {\bf k'}{\cdot}{\bbox{\rho}'})],
\]

\noindent
has the equation

\begin{equation}
\left[
-i\omega\tau - ik_d\rho_x + 1
\right]
{\sf {\cal R}}_{{\bbox{\rho}} {\bbox{\rho}'}}(\omega)
= \tau \delta({\bbox{\rho}} - {\bbox{\rho}'}) +
{ { {\sf {\cal R}}_{{\bf 0} {\bbox{\rho}'}} (\omega) }
  \over
  { F^{\rm eq}_0
} }
F^{\rm eq}_{\rho}.
\label{apxT3}
\end{equation}

\noindent
For ${\bbox{\rho}} = {\bf 0}$ this leads to

\begin{equation}
{\sf {\cal R}}_{{\bf 0} {\bbox{\rho}'}}(\omega)
= -{ {\delta({\bbox{\rho}'}) } \over { i(\omega + i\eta)} }.
\label{apxT4}
\end{equation}

\noindent
On the other hand, the low-frequency adiabatic part of 
${\sf {\cal R}}_{{\bbox{\rho}} {\bbox{\rho}'}}$
scales with the steady-state solution $F_{\bbox{\rho}}$
[in a collision-time model the asymptotic form
$F_{\bbox \rho}/{1\over 2}{n}$ replaces
${\Delta F}_{\bbox \rho}/{1\over2}{\Delta {n}}$].
On denoting the correlated part by 
${\sf {\cal C}}_{{\bbox{\rho}} {\bbox{\rho}'}}$
and recalling that the adiabatic part
exhausts the normalization of
${\sf {\cal R}}_{{\bf 0} {\bbox{\rho}'}}$,
we obtain

\begin{equation}
{\sf {\cal R}}_{{\bbox{\rho}} {\bbox{\rho}'}}(\omega) =
{\sf {\cal C}}_{{\bbox{\rho}} {\bbox{\rho}'}}(\omega)
- { {\delta({\bbox{\rho}'})}\over {i(\omega + i\eta)} }
{ {F_{\bbox{\rho}}}\over {F_{\bf 0}} }.
\label{apxT5}
\end{equation}

\noindent
When the above is put together with
equations (\ref{apxT2})--(\ref{apxT4}) we arrive,
after some algebra, at the explicit formula
for the correlated propagator:

\begin{equation}
{\sf {\cal C}}_{{\bbox{\rho}} {\bbox{\rho}'}}(\omega) =
\tau { { \delta({\bbox{\rho}} - {\bbox{\rho}'})
     - {\displaystyle { {F_{\bbox{\rho}}}\over
			{F_{\bf 0}} } }
	 \delta({\bbox{\rho}'}) }
 \over
{ 1 - i k_d \rho_x - i \omega \tau } }.
\label{apxT6}
\end{equation}

We can use equation (\ref{apxT6}) directly to evaluate both dissipative and
non-dissipative contributions to the noise. Using
the reciprocal-space representation
${\bf v} \leftrightarrow -i(\hbar/m^*)\partial/\partial {\bbox \rho}$,
the power density $P$ of equation (\ref{AY16}) is

\begin{eqnarray}
P
%IOP+ &=
=&&
2 { {e^2 E^2}\over {k_{\rm B} T} }
{\left( -{i\hbar\over m^*} \right)}^2
{ \left\{
{ {\partial}\over {\partial \rho_x} }
\int {d^{\nu} \rho'}
{\sf {\cal C}}_{{\bbox \rho} {\bbox \rho'}}(0)
{ {\partial}\over {\partial \rho'_x} }
\Delta F^{\rm eq}_{\rho'}
\right\} }_{\rho \to 0}
\cr
{\left. \right.} \cr
%IOP+ &=
=&&
2 { {e^2 E^2 \tau}\over {k_{\rm B} T} }
{\left( {\hbar\over m^*} \right)}^2
{ \left\{
-{ {\partial}^2\over {\partial \rho_x^2} }
\Delta F^{\rm eq}_{\rho}
\right\} }_{\rho \to 0}
\cr
{\left. \right.} \cr
%IOP+ &=
=&&
\sigma E^2.
\label{apxT8}
\end{eqnarray}

\noindent
The Drude conductivity $\sigma = {n} e \mu_{\rm e}$
appears when we apply the relation

\[
{\left\{
-{ {\partial^2}\over
   {\partial \rho_x^2}} \Delta F^{\rm eq}_{\rho}
\right\}}_{\rho \to 0}
= {\langle k^2_x \Delta f^{\rm eq} \rangle}
= {{m^*k_{\rm B}T}\over \hbar^2} {\left( {{n}\over 2} \right)}
\]

\noindent
to the middle line of the equation.
A contribution containing
$\langle v_x \Delta f^{\rm eq} \rangle = 0$
vanishes trivially.

The hot-electron spectral density
$S_g$ in the static limit
[recall equation (\ref{eq22b})]
is calculated similarly:

\begin{eqnarray}
%IOP+ \fl
S_g
%IOP+ &=
&&=
2{ {(e{\bf E}{\bbox \cdot}{\bf {\hat x}})^3}\over \hbar}
{ \left\{
\int {d^{\nu} \rho'} \int {d^{\nu} \rho''}
v_x {\sf {\cal C}}_{{\bbox \rho} {\bbox \rho'}}(0)
v'_x {\sf {\cal C}}_{{\bbox \rho'} {\bbox \rho''}}(0)
(-i\rho''_x \Delta F^{\rm eq}_{\rho''})
\right\} }_{\rho \to 0}
\cr
{\left. \right.} \cr
%IOP+ &=
&&=
2{ {e^3 E^3 \tau^2 \hbar}\over {m^*}^2 }
{\Biggl\{
  {\left[
    { {\partial}\over {\partial \rho_x} }
	     { 1\over {1 - ik_d \rho_x} }
    {\left(
      { {\partial}\over {\partial \rho_x} }
      { {-i\rho_x \Delta F^{\rm eq}_{\rho}}\over
	{1 - ik_d \rho_x} }
    \right)}
  \right]}_{\rho \to 0}
\Biggr.}
\cr
%IOP+ &{~~~~}
&&{~~~~}
{\Biggl. - {\left[
{ {\partial}\over {\partial \rho_x} }
{ {F_{\bbox \rho}/F_{\bf 0}}\over
  {1 - ik_d \rho_x} } \right]}_{\rho \to 0}
   {\left[
{ {\partial}\over {\partial \rho'_x} }
{ {-i\rho'_x \Delta F^{\rm eq}_{\rho'}}\over
   {1 - ik_d \rho'_x} } \right]}_{\rho' \to 0}
\Biggr\} }.
\label{apxT9}
\end{eqnarray}

\noindent
We evaluate this with the help of the relations
$\Delta F^{\rm eq}_0 = {1\over 2}\Delta {n}$ and
$\{ \partial F_{\bbox \rho}/\partial \rho_x \}_{\rho \to 0}
= ik_dF_{\bf 0}$, the latter following from equation (\ref{apxT2}).
The result is

\begin{equation}
S_g
= \sigma m^* \mu_{\rm e}^2 E^4
{\left( {{\Delta {n}}\over {n}} \right)}.
\label{apxT10}
\end{equation}

%-> 

%IOP+ \end{thebibliography}

%\newpage
%\section*{FIGURE CAPTIONS}
%IOP+ \Figures

\begin{figure}
\caption{
Zero-frequency spectral density of the excess (hot-electron)
thermal noise, above the equilibrium noise,
in a degenerate and uniform two-dimensional electron gas.
The excess is plotted as its ratio with
the zero-field noise ${\cal S}(E\!=\!0) = 4{\cal G}k_{\rm B}T$
as a function of driving field $E$ and for temperatures $T$
between 0 and 900 K, in steps of 150 K.
The dot-dashed line is at $T = 300$ K.
In the degenerate limit $T \ll T_{\rm F}$,
thermal noise scales with $T$;
thus the excess-noise ratio is independent of temperature.
At high temperature, the excess ratio for a given field value
diminishes as its denominator ${\cal S}(0)$ becomes dominant.
}
%\label{fig1}
\end{figure}


\begin{references}
%IOP+ \section*{References}

%IOP+ \begin{thebibliography}{38}

%01
\bibitem{liefr}
Liefrink F, Dijkhuis J I, de Jong M J M, Molenkamp L W 
and van Houten H 1994 {\it Phys. Rev.} B {\bf 49} 14066 

%02
\bibitem{rez}
Reznikov M, Heiblum M, Shtrikman H and Mahalu D 1995
{\it Phys. Rev. Lett.} {\bf 75} 3340

%03
\bibitem{ksgje}
Kumar A, Saminadayar L, Glattli D C, Jin Y and Etienne B
1996 {\it Phys. Rev. Lett.} {\bf 76} 2778

%04
\bibitem{smd}
Steinbach A H, Martinis J M and Devoret M H 1996
{\it Phys. Rev. Lett.} {\bf 76} 3806

%05
\bibitem{sbkpr}
Schoelkopf R J, Burke P J, Kozhevnikov A A, Prober D E
and Rooks M J 1997 {\it Phys. Rev. Lett.} {\bf 78} 3370

%06
\bibitem{poth}
Pothier H, Gu\'eron S, Birge N O, Esteve D and Devoret M H
1997 {\it Phys. Rev. Lett.} {\bf 79} 3490

%07
\bibitem{depic}
de Picciotto R, Reznikov M, Heiblum M, Umansky V,
Bunin G and Mahalu D 1997 {\it Nature} {\bf 389} 162

%08
\bibitem{sami}
Saminadayar L, Glattli D C, Jin Y and Etienne B 1997
{\it Phys. Rev. Lett.} {\bf 79} 2526

%09
\bibitem{theor}
de Jong  M J M and Beenakker C W J 1997
{\it Mesoscopic Electron Transport (NATO ASI Series E)}
ed L P Kouwenhoven, G Sch\"on and L L Sohn 
(Kluwer Academic, Dordrecht)

\bibitem{khlus}
Khlus V A 1987 {\it Sov. Phys. JETP} {\bf 66} 1243

\bibitem{lesovik}
Lesovik G B 1989 {\it JETP Lett.} {\bf 49} 592

%11
\bibitem{but}
B\"uttiker M 1992 {\it Phys. Rev. Lett.} {\bf 65} 2901;
1992 {\it Phys. Rev.} B {\bf 46} 12485

%12
\bibitem{beebut}
Beenakker C W J and B\"uttiker M 1992
{\it Phys. Rev.} B {\bf 46} 189

%13
\bibitem{thmldr}
Martin Th and Landauer R 1992 {\it Phys. Rev.} B {\bf 45} 1742

%14
\bibitem{nagaev1}
Nagaev K E 1992 {\it Phys. Lett.} A {\bf 169} 103;
1995 {\it Phys. Rev.} B {\bf 52} 4740

%15
\bibitem{djb2}
de Jong M J M and Beenakker C W J 1995
{\it Phys. Rev.} B {\bf 51} 16867

\bibitem{blbu}
Blanter Ya and B\"uttiker M
1999 cond-mat/9910158.

%__
\bibitem{datta}
Datta S 1995 {\it Electronic Transport in Mesoscopic Systems}
(Cambridge University Press, Cambridge)

%--
\bibitem{csr}
For example: the compressibility sum rule (see reference
\onlinecite{pinoz} below) links the magnitude of electron-hole
correlations to the total density of a degenerate system. In the
diffusive approaches, the correlations are perforce keyed
to the notional density of mobile diffusers {\it only},
because the carriers deep in the Fermi sea are excluded
from the calculation of diffusive transport
\cite{datta}
(moreover, this notional density is defined {\it ad hoc}).
This leads directly to an unphysical screening response
and violation of quasi-neutrality (perfect screening sum rule)
over the size of the sample.

%16
\bibitem{upon99}
Green F and Das M P 2000 {\it Proceedings of the Second
International Conference on Unsolved Problems of Noise
and Fluctuations (UPoN'99)} ed D Abbott and L B Kish
AIP {\bf 511} (American Institute of Physics, New York)
pp 422-33. For a similar discussion see cond-mat/9905086.
 
%17
\bibitem{ithaca}
Das M P and Green F 1999 in {\it Proceedings of the 23rd International
Workshop on Condensed Matter Theories} ed G S Anagnostatos
(Nova Science, in preparation); see also cond-mat/9910183

\bibitem{lecce1}
Reggiani L, Reklaitis A, Gonz\'alez T, Mateos J,
Pardo D and Bulashenko O M 2000 {\it Aust. J. Phys.} {\bf 53} 3

%19
\bibitem{mc}
Tadyszak P, Danneville F, Cappy A, Reggiani L,
Varani L and Rota L 1996 {\it Appl. Phys. Lett.} {\bf 69} 1450

\bibitem{lecce2}
Gonz\'alez T, Mateos J, Pardo D, Varani L and Reggiani L
1999 cond-mat/9910125

\bibitem{kormay}
Korman C E and Mayergoyz I D 1996 {\it Phys. Rev.} B {\bf 54} 17620

%24
\bibitem{sw2}
Stanton C J and Wilkins J W
1987 {\it Phys. Rev.} B {\bf 35} 9722;
1987 {\it Phys. Rev.} B {\bf 36} 1686

%25
\bibitem{sw0} 
Stanton C J 1986 Ph.D. thesis
(Cornell University, unpublished)

%18
\bibitem{ferry}
Ferry D K and Goodnick S M
1997 {\it Transport in Nanostructures}
(Cambridge University Press, Cambridge, UK)

%\bibitem{roukes}
%Roukes M L, Freeman M R, Germain R S, Richardson R C and Ketchen M B
%1985 {\it Phys. Rev. Lett.} {\bf 55} 422

\bibitem{gdii}
Green F and Das M P 2000 submitted to {\it J. Phys. Condensed Matter}.
See also cond-mat/9911251

%20
\bibitem{gdcond}
For a preliminary account see
Green F and Das M P 1998 cond-mat/9809339
(CSIRO-RPP3911, unpublished).

\bibitem{kogancpu}
Kogan Sh M 1996 {\it Electronic Noise and Fluctuations in Solids}
(Cambridge University Press, Cambridge, UK)

%21
\bibitem{pinoz}
Pines D and Nozi\`eres P
1966 {\it The Theory of Quantum Liquids}
(Benjamin, New York)

%22
\bibitem{kogan}
Kogan Sh M and Shul'man A Ya 1969
{\it Zh. Eksp. Teor. Fiz.} {\bf 56} 862
[1969 {\it Sov. Phys. JETP} {\bf 29} 467]

%23
\bibitem{ggk}
Gantsevich S V, Gurevich V L and Katilius R
1979 {\it Nuovo Cimento} {\bf 2} 1

%26
\bibitem{gc}
Green F and Chivers M J
1996 {\it Phys. Rev.} B {\bf 54} 5791

%27
\bibitem{nvk}
van Kampen N G 1981 {\it Stochastic Processes in Physics
and Chemistry} (North-Holland, Amsterdam) pp 246-52

%28
\bibitem{embed}
The Poisson equation is always defined in three dimensions.
To interpret equation (\ref{AX1.2a}) appropriately when $\nu < 3$,
the electron density ${\langle f \rangle}$ must be
understood to carry a (separable) factor in the $3 - \nu$
transverse space co-ordinates. Thus, for a transport problem
confined strictly to two dimensions, the Poisson source term contains
${\langle f({\bf r}) \rangle}
\equiv \delta(z) {\langle f({\bf r}_{\perp}) \rangle}$,
where $z$ is orthogonal to the plane $({\bf r}_{\perp}; z = 0)$.
On the other hand, the stabilizing background distribution
$n^+({\bf r})$ can be fully three-dimensional,
as in a modulation-doped heterostructure
\cite{ferry}.

%_%
\bibitem{variat}
The subtraction of left- and right-hand sides of equation
(\ref{AX1.1}) from (\ref{AX1}) is formally necessary,
despite the fact that both are identically zero. This is because
we will later require their functional derivatives with respect
to $f^{\rm eq}$. Those variations do {\it not} vanish identically,
as a quick test on equation (\ref{AX1.1}) shows.


%29
\bibitem{wmws}
We follow convention in taking the external field as primitive.
It induces the system response to be described by transport theory,
but is not itself describable at that level.
A more complete way of incorporating
electromotive forces into the physics of transport is given in
Magnus W and Schoenmaker W
1998 {\it J. Math. Phys.} {\bf 39} 6715.

%30
\bibitem{fg1}
Green F 1996 {\it Phys. Rev.} B {\bf 54} 4394

\bibitem{kb}
Kadanoff L P and Baym G 1962 {\it Quantum Statistical Mechanics}
(W A Benjamin, Reading, Massachusetts)
%31
\bibitem{fg1warn}
Note: a remark in reference \cite{fg1},
that $\Delta f(t)$ is inherently time-dependent, holds
only for collision-time approximations.

%32
\bibitem{vvt}
Van Vliet C M 1994
{\it IEEE Trans. Electron Devices} {\bf 41} 1902

%33
\bibitem{nougier}
Nougier J P 1980
{\it Physics of Nonlinear Transport in Semiconductors}
ed D K Ferry, J R Barker, and C Jacoboni
(Plenum, New York) p 415 ff
% 1994 {\it IEEE Trans. Electron Devices} {\bf 41} 2034

%34
\bibitem{nerlul}
Reggiani L, Lugli P and Mitin V
1988 {\it Phys. Rev. Lett.} {\bf 60} 736

%36
\bibitem{mbix}
Green F and Das M P 1998
{\it Recent Progress in Many-Body Theories}
ed D Neilson and R F Bishop (World Scientific, Singapore) p. 102.
See also cond-mat/9709142

%35
\bibitem{landauer}
Landauer R 1993 {\it Phys. Rev.} B {\bf 47} 16427

%%37
\bibitem{bakshi}
Bakshi P M and Gross E P 1968 {\it Ann. Phys.} {\bf 24} 419

%%38
\bibitem{kubo}
Kubo R, Toda M and Hashitsume N 1991
{\it Statistical Physics II: Nonequilibrium Statistical Mechanics}
(2nd ed., Springer, Berlin) pp 199 and 200

%-> 
\end{references}
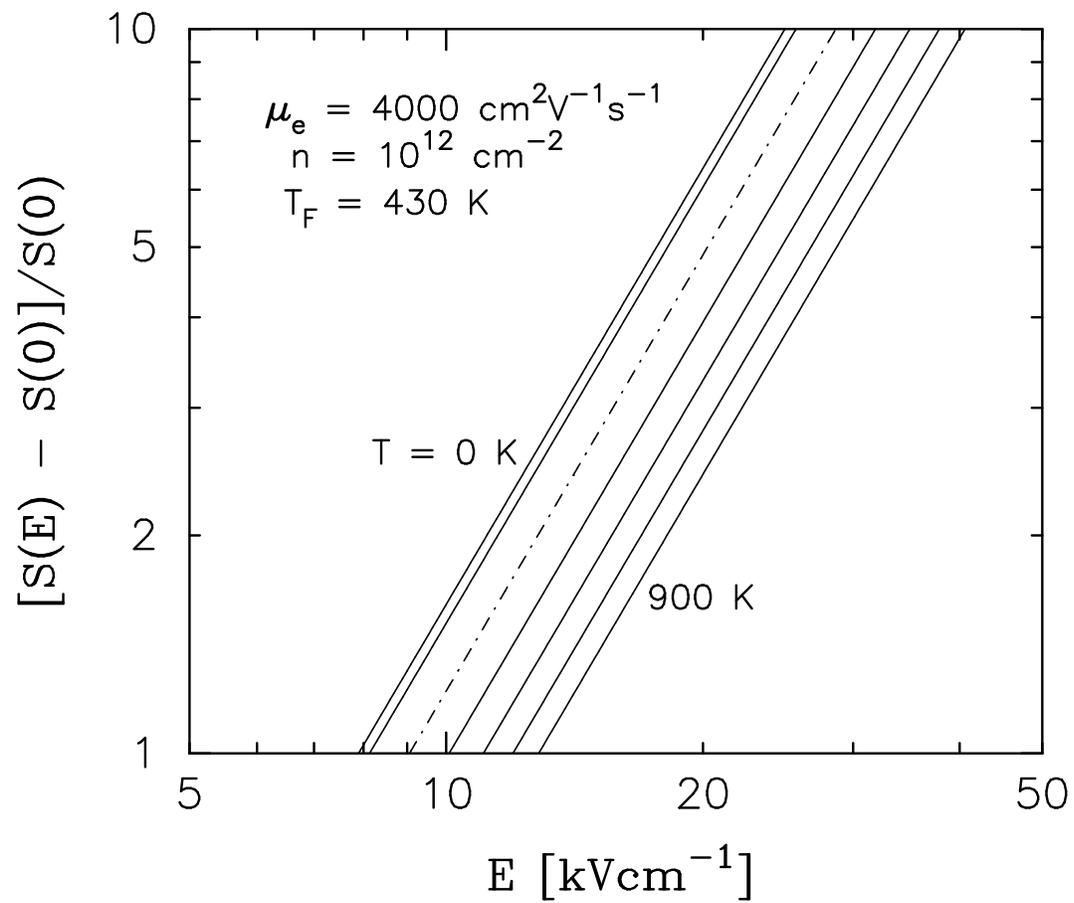
\end{document}